\documentstyle[aps,prd,t1enc,amsmath]{revtex}

\newcommand{\hs}{\,-\,}

\begin{document}

\title{Non\hs linear gravitational wave interactions with plasmas}

\author{Gert Brodin$^1$\footnote{E\hs mail:
    gert.brodin@physics.umu.se},
  Mattias Marklund$^{1,2,3}$\footnote{E\hs mail:
    mattias.marklund@physics.umu.se}
  and Peter K.\ S.\
  Dunsby$^3$\footnote{E\hs mail: peter@vishnu.mth.uct.ac.za}}

\address{$1$ Department of Plasma Physics,
  Ume{\aa} University, SE--901 87 Ume{\aa}, Sweden}
\address{$2$ Department of Electromagnetics, Chalmers University of
  Technology, SE--412 96 G\"oteborg, Sweden}
\address{$3$ Department of Mathematics and Applied Mathematics,
  University of Cape Town, Rondebosch 7701, South Africa}

\date{\today}

\maketitle

\begin{abstract}
We consider the interactions of a strong gravitational wave with
electromagnetic fields using the 1+3 orthonormal tetrad formalism.
A general system of equations are derived, describing the influence of
a plane fronted parallel (pp) gravitational wave on a cold relativistic
multi\hs component plasma. We focus our attention on phenomena that
are induced by terms that are higher order in the gravitational wave
amplitude. In particular, it is shown that parametric excitations of
plasma oscillations takes place, due to higher order gravitational
non\hs linearities. The implications of the results are discussed.
\end{abstract}

\draft

%%%%%%%%%%%%%%%%%%%%%%%%%%%%%%%%%%
\section{Introduction}
%%%%%%%%%%%%%%%%%%%%%%%%%%%%%%%%%%

There have been numerous investigations on the scattering of
electromagnetic waves off gravitational fields (see Refs.\
\cite{bi:Collection,bi:Macdonald-Thorne}). Previous research has
mostly directed
its interest towards the effects on vacuum electromagnetic fields
(although there are exceptions, see e.g.\ Refs.\
\cite{bi:Marklund-Brodin-Dunsby,bi:rest}, where the effects of plasmas
have been taken into account). Similarly, much work concerning
gravitational waves have
considered the linearized theory, which is obviously the relevant
regime for gravitational wave detectors, or, in general, for distances far
away from the gravitational wave source. Alternatively, there has been an
interest in exact solutions, and thus a number of exact gravitational
wave solutions (see e.g.\ Ref.\ \cite{bi:Bicak} and references therein)
have been found. In the present paper we will choose
an intermediate approach, starting with an exact gravitational wave
solution, but focusing on a weak amplitude (but still non\hs linear)
approximation, and studying the effects induced in a plasma.

The question under study in this paper is whether non\hs linear
gravitational wave effects -- that may be of significance close to the
gravitational wave source -- can give rise to qualitatively new
phenomena in plasmas that are absent in linearized theory. Close to 
the source, additional effects apart from non\hs linearities -- due to 
for example the three dimensional geometry and/or the non\hs radiative 
part of the gravitational field -- are likely to be important 
for astrophysical applications. However,
in order to focus on the processes directly induced by non\hs
linearities, a somewhat simpler model problem with a unidirectional 
gravitational wave will be studied:  To facilitate the
analysis of the non\hs linear interaction between a plasma and a
gravitational wave, we make use of the pp\hs wave solution of
Einstein's field equations. Furthermore, we introduce a Lorentz tetrad
in order to define physical variables in a straightforward manner. With
this setup, the governing plasma equations can be written in a simple
3\hs dimensional form. In Maxwell's equations, the gravitational effects
are given by effective charge- and current densities. Moreover, the fluid
equations are given for a cold plasma.

Previously, parametric excitation of a Langmuir wave and an
electromagnetic wave by a linearized gravitational wave has been 
considered \cite{bi:Brodin-Marklund}. Here we address the question
whether higher order terms in the gravitational wave amplitude can
result in new effects, using the above mentioned equations for a 
cold plasma. In order to demonstrate the usefulness of our set of
equations, we study the stability properties of a plasma in the
presence of a pp\thinspace -\thinspace\ wave.
We show that including $2^{{\rm nd}}$ order gravitational wave effects
may give rise to new phenomena. In particular it is found that
electrostatic waves can be excited at a resonant surface where
the gravitational wave frequency is equal to the local plasma
frequency. Our results are summarized and discussed in the last section
of the paper.
%%%%%%%%%%%%%%%%%%%%%%%%%%%%%%%%%%%%%
\section{Preliminaries}
%%%%%%%%%%%%%%%%%%%%%%%%%%%%%%%%%%%%%
\subsection{Equations for a general space-time}
%%%%%%%%%%%%%%%%%%%%%%%%%%%%%%%%%%%%%%%%%%%%%%%

We follow the approach presented in
\cite{bi:Marklund-Brodin-Dunsby} for
handling gravitational effects in Maxwell's equations. Suppose an
observer moves with 4-velocity $u^a$ ($a = 0, ..., 3$). This
observer will measure the electric and magnetic fields
\cite{bi:cargese}
\begin{equation}\label{split}
  E_a \equiv F_{ab}u^b \ , \quad
  B_a \equiv \tfrac{1}{2}\,\epsilon_{abc}F^{bc} \ ,
\end{equation}
respectively, where $F_{ab}$ is the EM field tensor. Here
$\epsilon_{abc}$ is the volume element on hyper\hs surfaces
orthogonal to $u^a$.

We denote the fluid velocity $V^a \equiv (\gamma,\gamma\boldsymbol{v})$,
where $\gamma \equiv (1 - \boldsymbol{v}^2)^{-1/2}$. Let $q$ be
the particle charge and $n$ the proper number density.
%%%%%%%%%%%%
Furthermore, we introduce the orthonormal frame (ONF) 
$\{{\boldsymbol{e}}_a, a = 0, ..., 4\}$, each of which is a linear
combination of the coordinate derivatives $\partial_i 
\equiv \partial/\partial x^i$, i.e., ${\boldsymbol{e}}_a 
= e_a\!^i\partial_i$.
%%%%%%%%%%%% 
Using the split (\ref{split}) together with $j^a = qnV^a$,
Maxwell's equations $\nabla_bF^{ab}
= j^a$, $\nabla_{[a}F_{bc]} = 0$ read
\begin{mathletters} \label{Maxwell}
\begin{eqnarray}
  {\boldsymbol{\nabla\cdot E}} &=& \rho_{\!_E} + \rho
  \label{eq:max1} \ , \\
  {\boldsymbol{\nabla\cdot B}} &=& \rho_{\!_B} \label{eq:max2}\ , \\
  \dot{\boldsymbol{E}} - {\boldsymbol{\nabla\times B}}
     &=& -{\boldsymbol{j}}_{\!_E} - {\boldsymbol{j}} \label{eq:max3}
     \ , \\
  \dot{\boldsymbol{B}} + {\boldsymbol{\nabla\times E}}
     &=& -{\boldsymbol{j}}_{\!_B}  \label{eq:max4} \ ,
\end{eqnarray}
\end{mathletters}
where the ``effective'' (gravity induced) charge densities and
current densities are
\begin{mathletters}\label{Masterbongo}
\begin{eqnarray}
  \rho_{\!_E} &\equiv& -\Gamma^{\alpha}\!_{\beta\alpha}E^{\beta}
  - \epsilon^{\alpha\beta\gamma}\Gamma^0\!_{\alpha\beta}B_{\gamma} \ ,
\label{eq:rhoe} \\
  \rho_{\!_B} &\equiv& -\Gamma^{\alpha}\!_{\beta\alpha}B^{\beta}
  + \epsilon^{\alpha\beta\gamma}\Gamma^0\!_{\alpha\beta}E_{\gamma} \ ,
\label{eq:rhob} \\
  j^{\alpha}_{\!_E} &\equiv& -(\Gamma^{\alpha}\!_{0\beta}
                       - \Gamma^{\alpha}\!_{\beta0})E^{\beta}
  + \Gamma^{\beta}\!_{0\beta}E^{\alpha}
  - \epsilon^{\alpha\beta\gamma}\left( \Gamma^0\!_{\beta0}B_{\gamma}
               + \Gamma^{\delta}\!_{\beta\gamma}B_{\delta} \right) \ ,
\label{eq:je} \\
  j^{\alpha}_{\!_B} &\equiv& -(\Gamma^{\alpha}\!_{0\beta}
                       - \Gamma^{\alpha}\!_{\beta0})B^{\beta}
  + \Gamma^{\beta}\!_{0\beta}B^{\alpha}
  + \epsilon^{\alpha\beta\gamma}\left( \Gamma^0\!_{\beta0}E_{\gamma}
               + \Gamma^{\delta}\!_{\beta\gamma}E_{\delta} \right) \ ,
\label{eq:jb}
\end{eqnarray}
\end{mathletters}
while $\rho \equiv \sum_{\rm p.s.}q\gamma n$ and ${\boldsymbol{j}}
\equiv \sum_{\rm p.s.}q\gamma n{\boldsymbol{v}}$ are the matter
charge and current densities, respectively (the sums are over all
particle species). 
%%%%%%%%%%%%%%%%
Here $\Gamma^a\!_{bc}$ are the Ricci rotation coefficients with 
respect to the ONF $\{{\boldsymbol{e}}_a\}$. We have introduced the
notation ${\boldsymbol{E}} \equiv (E^{\alpha}) = (E^1,
E^2, E^3)$ etc., ${\boldsymbol{\nabla}} \equiv ({\boldsymbol{e}}_1,
{\boldsymbol{e}}_2, {\boldsymbol{e}}_3)$, and the overdot stands for
derivative in the direction of the time-like vector 
${\boldsymbol{e}}_0$. 
%%%%%%%%%%%%%%%%%
The dot- and cross\hs products are defined in the usual Euclidean way. 

The energy\hs momentum tensor for each particle species is assumed to
take the form of pressure free matter (dust),  $T^{ab} = mnV^aV^b$, 
where $m$ is the rest mass of the particles.
Then the conservation equations $\nabla_bT^{ab} = qnF^{ab}V_b$ give
\begin{mathletters}\label{conservation}
  \begin{eqnarray}
    {\boldsymbol{e}}_0(\gamma n) + {\boldsymbol{\nabla\cdot}}(\gamma
    n{\boldsymbol{v}}) &=& -\gamma n\left( \Gamma^{\alpha}\!_{0\alpha}
    + \Gamma^{\alpha}\!_{00}v_{\alpha} +
    \Gamma^{\alpha}\!_{\beta\alpha}v^{\beta} \right) \ , \\
    \left( {\boldsymbol{e}}_0 + {\boldsymbol{v\cdot\nabla}}
    \right)\gamma{\boldsymbol{v}} &=& \frac{q}{m}({\boldsymbol{E}}
    + {\boldsymbol{v \times B}})
    - \gamma\left[ \Gamma^{\alpha}\!_{00} +
    \left( \Gamma^{\alpha}\!_{0\beta} + \Gamma^{\alpha}\!_{\beta 0}
    \right)v^{\beta} +
    \Gamma^{\alpha}\!_{\beta\gamma}v^{\beta}v^{\gamma}
    \right]{\boldsymbol{e}}_{\alpha}  \ .
  \end{eqnarray}
\end{mathletters}
%%%%%%%%%%%%%%%%%%%%%%%%%%%%%%%%%%%%%%%%%%%%%%%%%%%%%%%%%%%%%%%%
\subsection{Basic relations in the field of a pp\hs wave}
%%%%%%%%%%%%%%%%%%%%%%%%%%%%%%%%%%%%%%%%%%%%%%%%%%%%%%%%%%%%%%%%
Previous examinations of interactions between gravitational
radiation and EM waves have focused on linearized gravitation. On
the other hand, one may suspect that there will be interesting
effects in the non\hs linear regime, not present to linear order.
Below we will show that this is indeed the case.

In order to address the issue of how strong gravitational
radiation may be involved in generation of EM waves, we look at
the \emph{plane fronted parallel} (pp) waves
(for a discussion, see Refs.\
\cite{bi:pp-wave}), in the special case of
a linearly polarized plane wave
\begin{equation}\label{eq:pp}
  {\rm d}s^2 = -{\rm d}t^2 + a(u)^2\,{\rm d}x^2 + b(u)^2\,{\rm d}y^2 +
  {\rm d}z^2  \ ,
\end{equation}
where $u = z - t$, and $a$ and $b$ satisfy $ab_{uu} + a_{uu}b =
0$, and the subscript $u$ denotes a derivative with respect to retarded
time. Note that we have chosen a vacuum geometry, i.e. we have 
omitted the influence of the plasma on the metric.

In order to make interpretations simple, we introduce the
canonical Lorentz frame
\begin{equation}\label{frame}
  {\boldsymbol{e}}_0 = \partial_t \ , \ {\boldsymbol{e}}_1 =
  a^{-1}\partial_x \ , \
  {\boldsymbol{e}}_2 = b^{-1}\partial_y \ , \ {\boldsymbol{e}}_3 =
  \partial_z \ .
\end{equation}
With this frame, the effective charge and current densities
(\ref{Masterbongo}) read
\begin{mathletters}\label{bongo}
\begin{eqnarray}
  \rho_{\!_E} &=& -(\ln ab)_uE^3 \ , \label{eq:e-charge} \\
  \rho_{\!_B} &=& -(\ln ab)_uB^3 \ , \label{eq:b-charge} \\
  {\boldsymbol{j}}_{\!_E} &=& -(\ln b)_u(E^1 -
  B^2)\,{\boldsymbol{e}}_1 - (\ln a)_u(E^2 + B^1)\,{\boldsymbol{e}}_2
  - (\ln ab)_uE^3\,{\boldsymbol{e}}_3   \ ,
  \label{eq:e-current} \\
  {\boldsymbol{j}}_{\!_B} &=& -(\ln b)_u(E^2 +
     B^1)\,{\boldsymbol{e}}_1 + (\ln a)_u(E^1 -
     B^2)\,{\boldsymbol{e}}_2 - (\ln ab)_uB^3\,{\boldsymbol{e}}_3  \
     . \label{eq:b-current}
\end{eqnarray}
\end{mathletters}

Apart from Maxwell's equations (\ref{eq:max1})--(\ref{eq:max4})
[together with the effective charge and current densities
(\ref{eq:e-charge})--(\ref{eq:b-current})] we also need the fluid
equations. From the conservation equations (\ref{conservation}) we
obtain, the fluid equations using the frame (\ref{frame}):
\begin{mathletters}\label{fluideqs}
\begin{eqnarray}
  \frac{\partial}{\partial t}(\gamma n)
    + {\boldsymbol{\nabla\cdot}}(\gamma n{\boldsymbol{v}})
  &=& \gamma n(\ln ab)_u(1 - v_{\parallel})
    \ , \label{energy} \\
  \left(\frac{\partial}{\partial t} + {\boldsymbol{v\cdot\nabla}}
  \right) \gamma{\boldsymbol{v}}
  &=& \frac{q}{m}({\boldsymbol{E}} + {\boldsymbol{v\times B}})
  + \gamma\left[ (\ln a)_uv_1{\boldsymbol{e}}_1 + (\ln
b)_uv_2{\boldsymbol{e}}_2 \right](1 - v_{\parallel})
  + \gamma\left[ (\ln a)_uv_1^2 +(\ln b)_uv_2^2
    \right]{\boldsymbol{e}}_3  \ , \label{momentum}
\end{eqnarray}
\end{mathletters}
where $v_{\parallel} \equiv v_3$ is the velocity parallel to the
gravitational wave propagation direction. These equations should be
satisfied for each particle species. 
%%%%%%%%%%%%%%%%%%%%%
In the limit of small gravitational wave amplitudes 
and non\hs relativistic velocities, 
Eqs.\ (\ref{bongo})--(\ref{fluideqs}), together with Maxwell's
equations, were given in Ref.\ \cite{bi:Marklund-Brodin-Dunsby}. 
%%%%%%%%%%%%%%%%%%%%%
All terms with factors $(\ln ab)_u$ are however new, and -- as we 
will demonstrate in the remainder of this article -- they may induce 
new phenomena, compared to the linear regime.

%%%%%%%%%%%%%%%%%%%%%%%%%%%%%%%%%%%%%%%%%%%%%%%%%%%%%%%%%%%%%%%%%%
\section{An example: Parametric excitation of plasma oscillations}
%%%%%%%%%%%%%%%%%%%%%%%%%%%%%%%%%%%%%%%%%%%%%%%%%%%%%%%%%%%%%%%%%%

The longitudinal ``currents'' and ``charges'' are second order in the 
gravitational wave amplitude (see Appendix for further details).
These second order terms can give rise to qualitatively new
phenomena compared to the linear regime, and we demonstrate this by 
considering a simple, but illustrative, example.
In what follows, we will investigate longitudinal perturbations,
i.e., ${\boldsymbol{E}} = (0, 0, E)$, ${\boldsymbol{v}} = (0, 0,
v)$ etc., around a cold one-component equilibrium plasma. 
Compared to the case of weak gravitational waves
\cite{bi:Marklund-Brodin-Dunsby}, we now have $\rho_{\!_{E,B}}$
different from zero, and we also have a longitudinal contribution to 
the effective currents. This means that longitudinal EM- and plasma
waves can be excited.

In the unperturbed plasma, $\partial n_0/\partial t = 0$,
${\boldsymbol{E}}_0 = \boldsymbol{0}$, and ${\boldsymbol{B}}_0 =
\boldsymbol{0}$ \cite{bi:magnetic}.
We denote the number density perturbation by $\bar{n}$, i.e., $n(z,t) =
n_0(z) + \bar{n}(z,t)$ and assume that all perturbed quantities only
depend on $t$ and $z$. To first order, Maxwell's equation
(\ref{eq:max3}) becomes
\begin{eqnarray}
  \frac{\partial E}{\partial t} &=& (\ln ab)_uE
  - \mu_0qn_0v  \ , \label{eq:electric}
\end{eqnarray}
where we have used $j_m = qn_0v$.
Furthermore, the momentum equation (\ref{momentum}) becomes
\begin{equation}\label{momentum2b}
  \frac{\partial v}{\partial t} = \frac{q}{m}E \ .
\end{equation}

Taking the time derivative of Eq.\ (\ref{eq:electric}) and using
Eq.\ (\ref{momentum2b}), we obtain
\begin{equation} \label{eq:plasmaosc}
  \frac{\partial^{2}E}{\partial t^2} + \omega_{\mathrm{p}}^2(z)E
  = \frac{\partial}{\partial t}\left[ (\ln ab)_uE \right] \ ,
\end{equation}
where $\omega _{{\rm p}}(z)=[n_{0}(z)q^{2}\mu _{0}/m]^{1/2}$ is
the local plasma frequency. Thus the left hand side is the usual
equation for plasma oscillations in a cold inhomogeneous plasma,
and the right hand side is the modification induced by the the
pp\hs wave.
We next focus ourselves on weak periodic deviations from flat
space-time (see the Appendix), where the periodicity is $2\pi/\omega$.
At the resonant surface where $\omega_{{\rm p}}(z_{{\rm res}}) =
\omega$, we can then have parametric excitation of plasma
oscillations. We let $E(z_{{\rm res}},t) = \hat{E}(t)\exp \left[-{\rm i}\omega
t\right] + {\rm c.c.}$, where ${\rm c.c.}$\ denotes the complex
conjugate, and assume that $|\partial \hat{E}(t)/\partial t|
\ll \omega | \hat{E}(t)|$. At the resonant surface Eq.\
(\ref{eq:plasmaosc}) then reduces to
\begin{equation}\label{eq:elampl}
  \frac{d\hat{E}}{dt}
  = -{\tfrac{1}{2}}{\rm i}\exp\left( 2{\rm i}\omega z_{{\rm res}}
\right)
  \omega\hat{h}^2\hat{E}^{*} \ ,
\end{equation}
where the star denotes complex conjugate (see the Appendix). 
Taking the time-derivative of Eq.\ (\ref{eq:elampl}) and using the complex
conjugate of the same equation, we find
$\hat{E} \propto \exp(\Gamma t)$ where the growth rate is
\begin{equation}
  \Gamma = {\tfrac{1}{2}}\omega|\hat{h}^2|\;.
\end{equation}
Note that the threshold value for excitation is zero, since we
have not included any dissipation mechanism of the plasma
oscillations. Adding electron--ion collision in Eq.\ (\ref{momentum}),
the threshold value $\hat{h}_{\rm thr}$ of this instability is of
the order $(\nu_{\rm e-i}/\omega_{\rm p})^{1/2}$, where $\nu_{\rm
  e-i}$ is the electron--ion collision frequency.

Clearly, our instability does not occur unless higher order gravitational
perturbations are included, in contrast to the results in Ref.\
\cite{bi:Brodin-Marklund}. Thus the
corresponding growth rate is smaller in our case for a given
source of gravitational radiation. There are still two interesting
properties of the above instability as compared to the process in
Ref.\ \cite{bi:Brodin-Marklund}, where parametric excitation of a
Langmuir wave and an electromagnetic wave was considered:

(i) The
frequency matching condition in our case is $\omega = \omega_{\rm p}$,
which requires a rather high gravitational frequency \cite{Comment},
but is less severe than the condition in Ref.\
\cite{bi:Brodin-Marklund}, where $\omega \ge 2\omega_{\rm p}$.

(ii) In
contrast to most parametric instabilities in plasmas we have no wave
vector matching condition, but instead the process takes place at a
localized resonance surface $z = z_{\rm res}$ where $\omega =
\omega_{\rm p}(z_{\rm res})$. This means that there is no threshold
value for the instability introduced by plasma
inhomogeneities. Normally the threshold value is inversely
proportional to the inhomogeneity scale length
\cite{bi:Rosenbluth-White-Liu}, and close to a binary system, where
the effects of gravitational radiation are likely to be most
important, such a condition for parametric excitation may thus be
rather severe. Unfortunately, the result of the ``no inhomogeneity
threshold'' depends on the cold plasma approximation, and a finite
temperature is likely to change the picture.

%%%%%%%%%%%%%%%%%%%%%%%%%%%%%%%%%%%%%%%
\section{Summary and discussion}
%%%%%%%%%%%%%%%%%%%%%%%%%%%%%%%%%%%%%%%

In the present paper we have investigated a higher order effect of
gravitational waves on a plasma. For this purpose we have developed a
Lorentz tetrad formalism for a cold
plasma in the presence of a strong gravitational wave. 
The Lorentz tetrad approach has of course been widely used before, 
perhaps most notably in the membrane paradigm approach to
black hole spacetimes \cite{bi:Macdonald-Thorne}.
The obvious advantage of using a Lorentz tetrad is its direct connection to
measurements. It is possible to formulate Maxwell's equations such
that the gravitational contributions takes the form of ``charge''- and
``current'' densities. Similarly, the fluid equations
are modified by effective particle sources and gravitational
forces. Of course, this is not the physical picture behind the
equations, but it still provides a useful tool for predicting the
consequences of the gravitational influence.

The main purposes of this study has been to (i) provide a framework
for investigating strong gravitational pulse effects in cold multi\hs
component plasmas, and (ii) to show that higher order
gravitational wave effects may be of importance, since they introduce
effective charges and longitudinal currents, as well as effective
``particle
sources'' and gravitational forces. As demonstrated, this in
turn makes new processes -- such as parametric generation of
electrostatic waves -- possible.
Since the effect under discussion is of order $\hat{h}^2$, we do
not believe that it will be of significance concerning {\em direct}
earth based observations of gravitational waves. It is
possible however that there exists favorable circumstances, e.g.\
close to a binary merger, for which the higher order
gravitational effective charge- and current densities can
play an important role. Close to such sources, the
gravitational wave amplitudes can reach considerable strength,
implying observational possibilities for the induced phenomena.

%%%%%%%%%%%%%%%%%%%%%%%%%%%%%%
\section*{Acknowledgments}
%%%%%%%%%%%%%%%%%%%%%%%%%%%%%%

M.\ M.\ was supported by the Royal Swedish Academy of Sciences. P.\
K.\ S.\ D.\ was supported by the NRC (South Africa).

%%%%%%%%%%%%
\appendix
%%%%%%%%%%%%%%%%%%%%%%%%%%%%%%%%%%%%%%%%%%%%%%%%%%%%%%%%%%%%%%%%
\section*{Perturbative expansion of the pp\hs wave}
%%%%%%%%%%%%%%%%%%%%%%%%%%%%%%%%%%%%%%%%%%%%%%%%%%%%%%%%%%%%%%%%

In many situations of interest the gravitational wave amplitude is
small,
i.e. $\left| a-1\right| \ll 1$, $\left| b-1\right| \ll 1$ and it is
appropriate to make approximations for the factors $(\ln a)_{u}$, $(\ln
b)_{u}$ and $(\ln ab)_{u}$ that appears in the gravitational source
terms in
Eqs.\ (\ref{bongo}) and (\ref{fluideqs}).
We will concentrate on approximately periodic
gravitational waves, such as those generated by binary systems, in order
to get definite results.
Let $a(u) = \sum_{n = -\infty}^{\infty}\hat{a}_n\exp({\rm i}n\omega u)$,
$b(u) = \sum_{n = -\infty}^{\infty}\hat{b}_n\exp({\rm i}n\omega
u)$, where $\hat{a}_n, \hat{b}_n \ll 1$,
$|\hat{a}_n| \sim |\hat{b}_n| \sim |\hat{a}_1|^{|n|}$, $\forall\
n$. Furthermore $\hat{a}_n^{\ast} = \hat{a}_{-n}$ and similarly for
$b$. Then $a_{uu}b + ab_{uu} = 0$ becomes
\begin{equation}\label{expansion}
  \sum_{n = -\infty}^{\infty}\sum_{m =
  -\infty}^{\infty}(n^2 + m^2)\hat{a}_n\hat{b}_m\exp[{\rm i}\omega(n +
  m)] = 0   \ .
\end{equation}
To $0^{\rm th}$  order, we assume that $\hat{a}_0 = 1 = \hat{b}_0$.
To first order, the solution to Eq.\ (\ref{expansion}) is $\hat{a}_1 =
-\hat{b}_1 \equiv \hat{h}$. Clearly, quadratic non\hs linear terms will
generate second harmonic terms proportional to $\exp(2{\rm i}\omega u)$.
Separating the frequencies in Eq.\ (\ref{expansion}),
and concentrating on the second harmonic part we obtain
\begin{equation}\label{eq:a2}
  2\hat{b}_2 + 2\hat{a}_2 - \hat{h}^2 = 0 \ .
\end{equation}
The canonical choice is $\hat{a}_2 = \hat{b}_2 =
(1/4)\hat{h}^2$.
Physically this means that we minimize the (pseudo)
energy density at the second harmonic frequency. Thus -- for this choice
-- all the oscillations at $2\omega$ are strictly due to the non\hs
linearity of Einstein's equations, and no harmonics are assumed to be initially
present, i.e.\ generated by a varying octopole moment of the binary
source. For astrophysical applications this is not necessarily the most
accurate choice (since binary systems may indeed have finite octopole
moments), but it has the advantage of clearly isolating the effects due
to non\hs linearities. Furthermore, it turns out that including the effect
of higher moments of the gravitational source (i.e.\ octupole moments and
higher) do not influence our calculations in Sec.\ III. The reason is
that an alternative solution to Eq.\ (\ref{eq:a2}) --
$\hat{a}_2 = (1/4)\hat{h}^2 + \delta\hat{a}_2$, $\hat{b}_{2} =
(1/4)\hat{h}^{2} - \delta\hat{a}_2$, instead of $\hat{b}_2 = \hat{a}_2
= (1/4)\hat{h}^2$ -- does not significantly affect the factor $\ln
(ab)_u$,
since it is independent of $\delta\hat{a}_2$ to second order in the
gravitational amplitude, provided $\delta\hat{a}_2 \sim \hat{h}^2$.

Continuing to $3^{\rm rd}$ order, a similar
calculation shows that we can make the natural choice
$\hat{a}_3 = \hat{b}_3 = 0$. For the $4^{\rm th}$ order terms,
Eq.\ (\ref{expansion}) gives
\begin{equation}
  32(\hat{a}_4 + \hat{b}_4) - \hat{h}^4 = 0 \ .
\end{equation}
where the canonical choice $\hat{a}_4 = \hat{b}_4 =
(1/64)\hat{h}^4$ is made. Continuing this procedure, it turns out
that all terms odd
in $n \neq 1$ disappear, while the terms even in $n$ satisfy
$\hat{a}_n = \hat{b}_n$ $\forall\ n$.

Using the above results, the logarithmic
factors in Eqs.\ (\ref{bongo}) and (\ref{fluideqs}) become
\begin{mathletters}
  \begin{eqnarray}
    (\ln a)_{u} &=& {\rm i}\omega \hat{h}\exp({\rm i}\omega u) -
    {\tfrac{1}{2}}{\rm i}\omega\hat{h}^2\exp(2{\rm i}\omega u)
    + {\tfrac{1}{4}}{\rm i}\omega\hat{h}^3\exp(3{\rm i}\omega u)
    - {\tfrac{1}{16}}\omega\hat{h}^4\exp(4{\rm i}\omega u)
    + {\rm c.c} \\
    (\ln b)_{u} &=& -{\rm i}\omega\hat{h}\exp({\rm i}\omega u) -
    {\tfrac{1}{2}}{\rm i}\omega\hat{h}^2\exp(2{\rm i}\omega u)
    - {\tfrac{1}{4}}{\rm i}\omega\hat{h}^3\exp(3{\rm i}\omega u)
    - {\tfrac{1}{16}}\omega\hat{h}^4\exp(4{\rm i}\omega u)
    + {\rm c.c} \\
    (\ln ab)_{u} &=& -{\rm i}\omega\hat{h}^2\exp(2{\rm i}\omega u)
    - {\tfrac{1}{8}}\omega\hat{h}^4\exp(4{\rm i}\omega u) + {\rm c.c}
\end{eqnarray}
\end{mathletters}
to $4^{\rm th}$ order in the gravitational amplitude.
This procedure may of course be continued to arbitrary order, noting
that this in general will result in an asymptotic series, i.e., it
does not {\em necessarily} converge towards a solution of $a_{uu}b +
ab_{uu} = 0$.


\begin{thebibliography}{99}
\bibitem{bi:magnetic} Note that as long as the magnetic field is
  parallel to the direction of propagation of the gravitational wave
  it will decouple from the rest of the equations, and our following
  results are therefore unaffected by
  the addition of such a field.
\bibitem{Comment} For interstellar matter we have $\omega_{\rm p} \sim
  10^4$ rad/s.
\end{thebibliography}
\end{document}